# Selection Rules for All-Optical Magnetic Recording in Iron Garnet

A. Stupakiewicz[1*], K. Szerenos[1,2], M. D. Davydova[3,4], K. A. Zvezdin[3,4], A. K. Zvezdin[3,4], A. Kirilyuk[2] and A. V. Kimel[2,5]

[1]Faculty of Physics, University of Bialystok, 15-245 Bialystok, Poland.
[2]Institute for Molecules and Materials, Radboud University, 6525 AJ, Nijmegen, The Netherlands.
[3]Moscow Institute of Physics and Technology (State University), 141700 Dolgoprudny, Russia.
[4]Prokhorov General Physics Institute, Russian Academy of Sciences, 119991 Moscow, Russia.
[5]Moscow Technological University (MIREA), 119454 Moscow, Russia.

**Finding an electronic transition a subtle excitation of which can launch dramatic changes of electric, optical or magnetic properties of media is one of the long-standing dreams in the field of photo-induced phase transitions [1-5]. Therefore the discovery of the magnetization switching only by a femtosecond laser pulse [6-10] triggered intense discussions about mechanisms responsible for these laser-induced changes. Here we report the experimentally revealed selection rules on polarization and wavelengths of ultrafast photo-magnetic recording in Co-doped garnet film and identify the workspace of the parameters (magnetic damping, wavelength and polarization of light) allowing this effect. The all-optical magnetic switching under both single pulse and multiple-pulse sequences can be achieved at room temperature, in narrow spectral ranges with light polarized either along <110> or <100> crystallographic axes of the garnet. The revealed selection rules indicate that the excitations responsible for the coupling of light to spins are *d*-electron transitions in octahedral and tetrahedral Co-sublattices, respectively.**

The most intriguing pathways to change the magnetization with a laser pulse are those that otherwise leave the entropy of the whole system almost intact. Such precisely directed impact has the potential to reduce energy waste. Recently it has indeed been shown that using a single 50 fs laser pulse one can permanently switch the magnetization in Co-doped yttrium iron garnet thin film (YIG:Co) [10] (see Fig. 1a). In this material, the Co dopants are responsible for strong magnetocrystalline anisotropy, making cube diagonals <111> to be eight easy directions. Optical excitation of localized *d-d* transition in Co ions at 0.95 eV with light polarized along [100] or [010] axis was able to modify the magnetic anisotropy and thus to switch the magnetization between the metastable states. This observation raised questions about the selection rules e.g. the feasibility of magnetic recording using light of other photon energies and light polarizations.

Aiming to define the workspace of the key parameters allowing the switching, we developed a phenomenological model of the photo-induced magnetization dynamics in YIG:Co film based on the Landau-Lifshitz-Gilbert (LLG) equation:

$$\dot{\boldsymbol{M}} = \gamma \left[\boldsymbol{M}, \frac{\delta \Phi}{\delta \boldsymbol{M}}\right] + \frac{\alpha}{M_S}[\boldsymbol{M}, \dot{\boldsymbol{M}}] + \boldsymbol{T}_E(\boldsymbol{M}, \boldsymbol{E}), \qquad (1)$$

where $\gamma$ is the gyromagnetic ratio, $\alpha$ is the Gilbert damping parameter, $M_S$ is the saturation magnetization, $E$ is the electric field vector, and $\Phi$ is the thermodynamic potential of the unperturbed magnet in the form of free energy.

The last term in the equation stands for optically-induced spin-torque $T_E(M, E)$. The latter may be represented in the following way in the Cartesian coordinate system:

$$T_E(M, E) = \gamma [M, H_{eff}(M)], \left(H_{eff}(M)\right)_i = \beta_{ijkl} M_j E_k E_l^* \qquad (2)$$

where $\beta_{ijkl}$ is the photo-magnetic tensor [10]. The number of independent tensor components can be found taking into account the *4mm* point group for (001)-oriented garnet film [11,12] and the fact that the tensor $\beta_{ijkl}$ must be invariant with respect to permutations of the last two indices. The material parameters and coefficients of the photo-magnetic tensor *a*, *b*, *c* (see Methods) were obtained by fitting the theoretical results to the experimental data from ref. 6.

There are four easy magnetization axes near <111>–type directions and 8 ground states for magnetization vector, taking into account two possible orientations along each axis. We denote these metastable magnetization states with numbers 1–8 (see in Fig. 1b). Earlier studies allowed to estimate that the life-time of the photo-magnetic anisotropy is about 20 ps [13]. The model accounts for this fact by introducing a temporal relaxation of the tensor components $\beta_{ijkl}$ accordingly. Therefore, the model predicts that during the first 20 ps after the optical excitation the motion of spins will mainly be defined by the field of the photo-induced magnetic anisotropy and afterwards the motion will be driven by the effective field of the steady-state magneto-crystalline anisotropy.

We demonstrate the modelling of the photo-magnetic switching between possible magnetization axes (phases), e.g. "1–4", "1–5", "1–8" (see Fig. 1b). Taking "1" as the initial magnetization state we performed the simulations of the trajectories of the magnetization vector varying the intensity of light and magnetic damping $\alpha$ (see Methods). Figure 1(c,d) summarizes the phase diagram of the simulations showing the final states to which the magnetization relaxes after being excited by light of two polarizations ($E\|[100]$ and $E\|[110]$). Fixing the damping at $\alpha=0.28$, which is very close to the experimentally observed value [10,13], the switching can be observed for both polarizations $E\|[100]$ and $E\|[110]$ in a wide range of light intensities. This model demonstrates different trajectories of the switching (see Fig. 1e) between: (i) only in-plane magnetization components (see "1–4" trajectory); (ii) only perpendicular magnetization components (see "1–5" trajectory); (iii) simultaneously in-plane and perpendicular magnetization components (see "1–8" trajectory).

Hence the simulations suggest that in addition to the photo-magnetic recording reported in ref. 10, the symmetry of the garnet films supports the possibility of the magnetic switching using other light polarizations than <100>. The mechanism of the switching must rely on excitation of other electronic transitions.

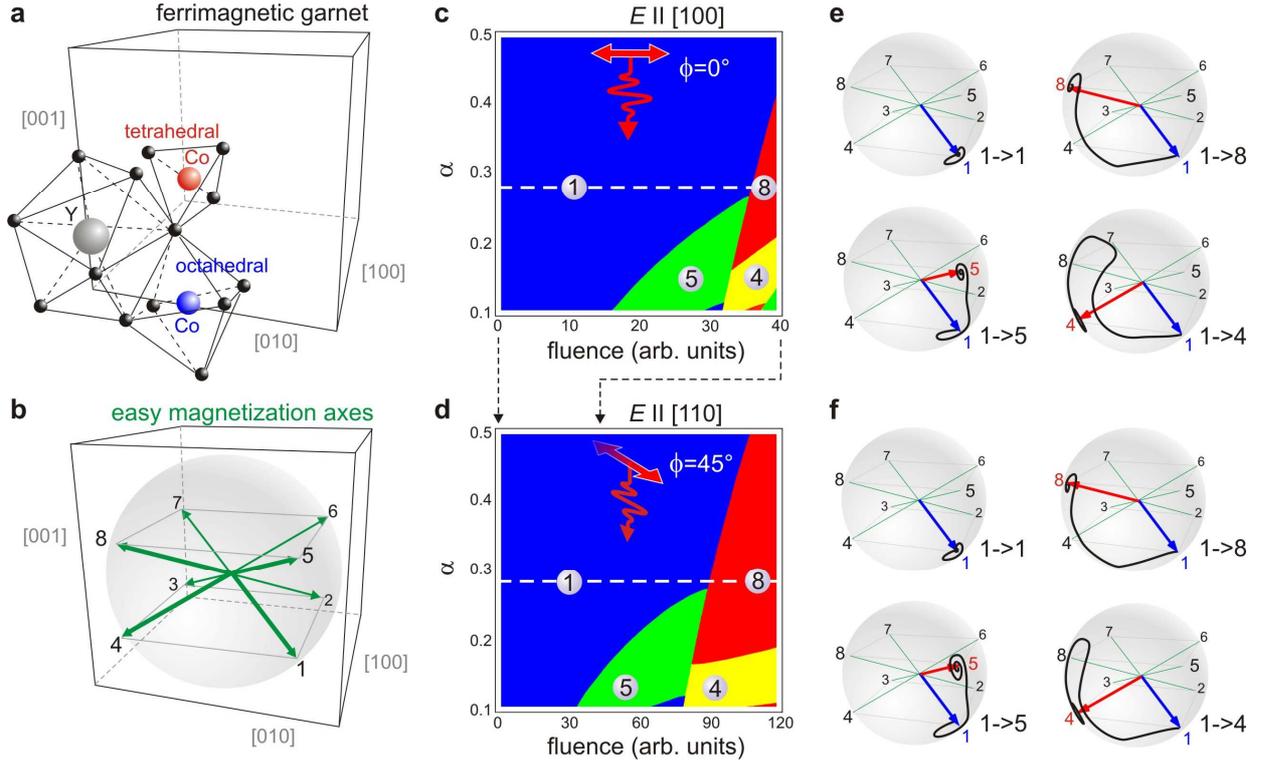

**Figure 1 | Photo-magnetic switching in YIG:Co**. (a) Crystal structure of iron-garnet. Co ions can enter the crystal by taking the positions in octahedral and tetrahedral environments. (b) Eight metastable states of the magnetization in relation to the crystal symmetry of iron-garnet. (c) Phase diagram of the photo-magnetic switching from state "1" with light polarization $E\|[100]$ as a function of Gilbert damping α and light fluence. (d) Phase diagram of the photo-magnetic switching from state "1" with light polarization $E\|[110]$ for as a function of Gilbert damping α and light fluence. (e) Trajectories of the switching from state "1" to states "4", "5", and "8" using light polarization $E\|[100]$. (f) Trajectories of the switching from state "1" to states "4", "5", and "8" using light polarization $E\|[110]$. In all shown trajectories the blue arrow is the initial magnetization state and the red arrow is the state after the switching. The calculations are performed for damping α=0.28. Fit parameters for all simulations are $b=-a/3$, $c=a/2$ (see Eq. 4 in Methods).

Compared to pure YIG, the absorption spectra of YIG:Co in the range of photon energies from 0.5 eV to 2 eV show a number of additional peaks associated with Co-ions [14]. The most pronounced peaks are observed around 1.13 eV (the wavelength is 1.1 μm) and 0.95 eV (the wavelength is 1.3 μm) [15-17]. So far the most conventional way to illustrate the switching is based on scanning a polarized laser beam, consisting of series of femtosecond laser pulses, across the sample surface [6-8]. To verify the feasibility of the magnetic switching at other photon energies and other polarizations of light, we employed this method for two polarizations of light [110] and [1-10]. The experiments were performed at several photon energies in the range from 0.83 eV to 1.24 eV.

The initial magnetic domain pattern as observed in a polarized light microscope (see Methods) is shown in Fig. 2. We rotated the linear polarization over 45° with respect to [100] axis ($\phi$ is the angle between the electric field of light and [100] axis) and searched for the wavelength allowing the switching. After scanning the beam with $\phi$ = 45° and photon energy 1.1 eV ($\lambda$ = 1140 nm) the large black domains are turned into white ones. The orthogonal polarization along [1-10] crystallographic axis does the opposite. Both polarizations are different from those used for the switching in ref. 10 indicating that the magnetic recording relies on a different microscopic mechanism which has not been reported before.

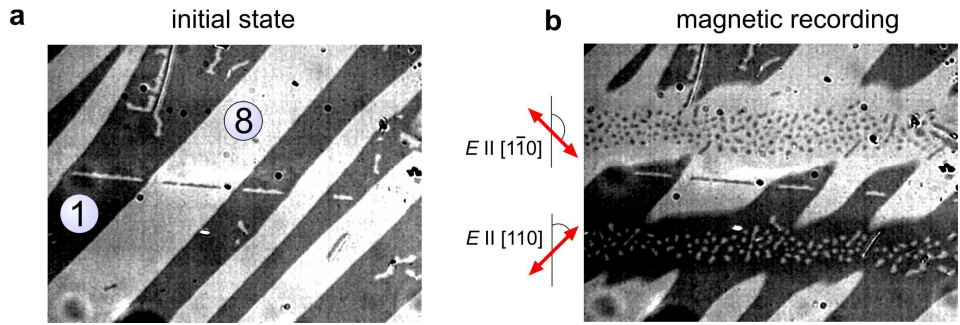

**Figure 2 | Magnetic domains recorded by scanning polarized laser beam.** (a) Initial domain pattern in YIG:Co film. Large black and white stripes correspond to the domains with two orientations of the magnetization – "1" and "8" magnetization states in Fig.1b. (b) The result of a scan with $E\|[110]$ polarized light ($\phi$ = 45°) and $E\|[1\text{-}10]$ polarized light ($\phi$ = 135°). The beam contains 50 fs laser pulses with the repetition rate of 1 kHz and the fluence of 100 mJ cm$^{-2}$. The central photon energy is 1.1 eV ($\lambda$ = 1140 nm). The recorded tracks are repeatable and stable for a long time at room temperature without external magnetic field. The images are 600×450 μm$^2$ large.

Figure 3 shows how the switched area increases with an increase of number of pump pulses *N*. (see Fig. 3a). By increasing the pump pulse fluence, the number of pulses required for the switching can be brought down to one. Note that the appearance of a domain after a certain number of pulses is a very reproducible effect that is determined by a balance between the repetition rate of the pulses and magnetostatic field driven domain wall motion (for more details see Methods and Extended Data Fig. 1).

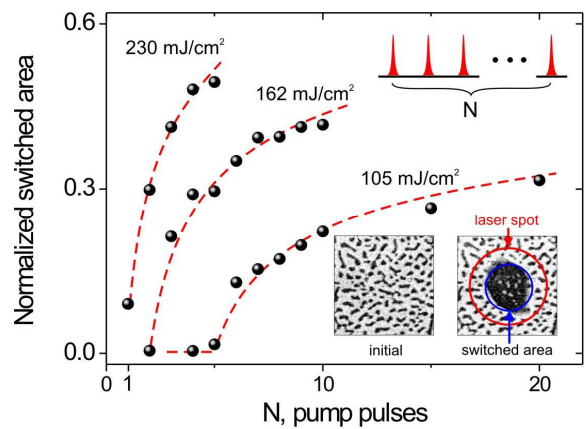

**Figure 3 | Photo-magnetic switching with multiple pulses at different pump fluences.** A single pulse with the fluence of 230 mJ/cm$^2$ is able to create a stable domain. The time period between the pulses is 1 ms (repetition rate was 1 kHz), the central photon energy of the pump is 1.1 eV ($\lambda$ = 1140 nm) and the pump polarization is along [110]. Dashed lines are guides to the eye. The inset shows the magnetic domain patterns before and after the laser excitation. The images are 200×200 μm$^2$ large.

Figure 4 shows a summary of the switching efficiencies as functions of light wavelength, polarization angle, and pulse fluence. The spectral dependence shown in Fig. 4(a) reveals two resonant features around 1.1 eV and 0.95 eV, respectively. Figure 4(b) shows that the most efficient photo-magnetic switching with photon energy of 0.95 eV is observed for [100] polarization. If the pumping is at photon energy of 1.1 eV, the switching has the maximum efficiency for [110] polarization. This exactly corresponds to the modelling results revealing different threshold intensities for the switching with different light polarizations (see phase diagrams in Fig. 1). Finally, pumping at 0.95 eV is characterized by 3 times lower threshold intensity compared to the case of pumping at 1.1 eV (Fig. 4c).

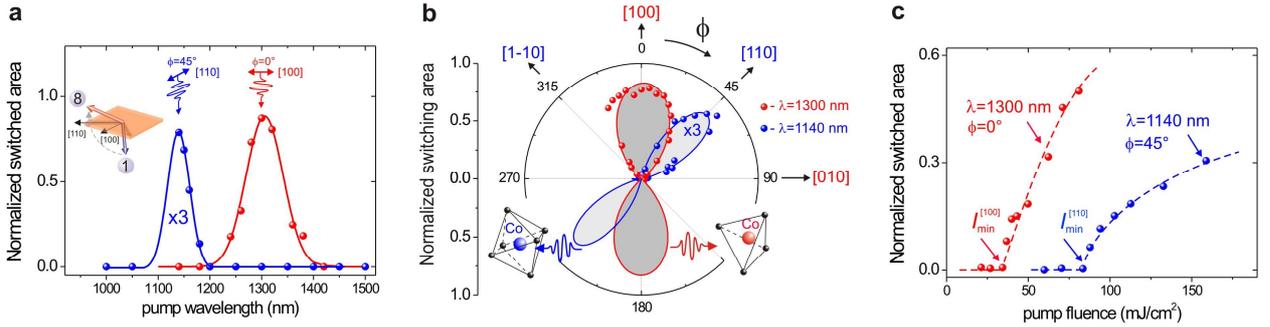

**Figure 4 | Selection rules of photo-magnetic switching on wavelength and polarization.** For measurements 5 laser pump pulses (N=5) were used. (a) Normalized switched area as a function of the photon energy for two polarizations $E\|[110]$ ($\phi = 45°$) and $E\|[100]$ ($\phi = 0°$). Solid lines were fitted by Gaussian function. (b) Normalized switched area as a function of the angle $\phi$ for the incoming polarization for two photon energies 0.95 eV ($\lambda$=1300 nm) and 1.1 eV (1140 nm). Solid lines are a fit by $cos(2\varphi)$ function. (c) Fluence dependence of the normalized switched area for two photon energies 0.95 eV and 1.1 eV. Dashed lines are guides to the eye. The measurements at different photon energies reveal different threshold fluences $I_{min}$.

The revealed selection rules shown in Figs. 4(a,b) allow us to identify the microscopic mechanisms responsible for the photo-magnetic recording. According to refs. [16,17] a sharp peak around 1.1 eV nm is due to $^4T_1 \rightarrow {^4T_2}$ electronic transition in $Co^{2+}$ ions in octahedral cites. A broader peak at 0.95 eV corresponds to $^5E \rightarrow {^5T_2}$ transitions in $Co^{3+}$ ions and $^4A_2 \rightarrow {^4T_1}$ transitions in $Co^{2+}$ ions in tetrahedral cites [16]. In both experiment and simulations the threshold fluence for the switching with light polarization along [110] axis is about three times larger than the threshold for light polarization along [100] axis (see Fig. 1(c,d) and Fig. 4c). The reason for this is possibly related to the fact that, as a general crystallographic property, the oscillator strength of low-symmetry tetrahedral sites is much stronger than high-symmetry octahedral sites [15,18,19]. As a result, the response of octahedral ions becomes clearly noticeable only when sufficient pump fluence is applied. The contribution to the magnetic anisotropy at room temperature is also stronger from Co-ions in the tetrahedral cites [20].

Finally, it is interesting to note the remarkable efficiency and selectivity of the switching by pumping different electronic transitions in Co-dopants. In the studied films only one of forty Fe ions is substituted by Co [21-23]. It means that single-photon excitation of Co ion must be, in principle, sufficient to control the spins in a rather large volume of 28 nm$^3$ (see Methods).

This work shows that tuning the wavelength, magnetic damping, polarization and fluence of light it is possible to find different regimes of the switching in a wide class of materials, which can possibly be even more efficient than those demonstrated before. The examples of suitable materials are ferrites (garnet, spinel, ortho- and hexaferrites, ferric borates, magnetite) [24], perovskites, spin glasses. For instance, a strong contribution to single ion magnetic anisotropy is typical for *3d* (e.g. $Mn^{2+}$, $Mn^{3+}$, $Cr^{2+}$, $Cr^{3+}$, $Fe^{2+}$, $Co^{2+}$, $Co^{3+}$, $Ni^{2+}$, $Ni^{3+}$), *4d* ($Ru^{3+}$, $Ru^{4+}$), *5d* ($Ir^{3+}$, $Ir^{4+}$) and *4f* ($Ce^{3+}$, $Tb^{3+}$) elements. Finding novel families of magnetic dielectrics switchable with laser pulses is crucial for future applications of photo-magnetism.

*Correspondence to: e-mail: and@uwb.edu.pl

**Acknowledgements** We acknowledge support from the project TEAM/2017-4/40 of the Foundation for Polish Science co-financed by the EU within the ERDFund, the Netherlands Organization for Scientific Research (NWO) and the programme 'Leading Scientist' of the Russian Ministry of Education and Science (14.z50.31.0034). We thank A. Maziewski and Th. Rasing for continuous support.

**Author Contributions** A.S. conceived the project with contributions from A.K. and A.V.K. The measurements were performed by K.S. The model was designed and the simulations were performed by M.D.D, K.A.Z, and A.K.Z. A.S., A.K., A.V.K. co-wrote the manuscript with contributions from K.S. and A.K.Z. The project was coordinated by A.S.

**METHODS**

**Materials**

The monocrystalline cobalt-doped yttrium iron garnet (YIG:Co) thin film was deposited on gadolinium-gallium garnet (GGG) substrate [22]. The miscut angle of the GGG substrate has been about 4°. The dopant $Co^{2+}$ and $Co^{3+}$ ions substitute $Fe^{3+}$ in both tetrahedral and octahedral sublattices, which are coupled antiferromagnetically. The Co ions enhance the magnetocrystalline anisotropy and result in a strong photo-magnetic effect, allowing to manipulate the magnetization with light [23]. In this sample the magnetic anisotropy constants both cubic ($K_1$= –8.4×10$^3$ erg cm$^{-3}$) and uniaxial ($K_U$= –2.5×10$^3$ erg cm$^{-3}$) were measured at room temperature. Incorporation of Co ions also increases the Gilbert damping parameter to a large value of $\alpha$=0.2. Both tetrahedral and octahedral $Fe^{3+}$ ions are substituted by $Ge^{4+}$ ions to decrease the saturation magnetization ($M_S$=7 G). $Ca^{2+}$ dopants enter the dodecahedral sites and take care of charge compensation. The Curie temperature was 455 K.

As single ion contribution to magnetic anisotropy from Co-ions is orders of magnitude higher than that of Fe-ions [20], a single photon can control spins in a large garnet volume. For YIG:Co film with the thickness of 7.5 um, the minimum pump intensity required for the switching was 34 mJ cm$^{-2}$. YIG:Co film is a ferrimagnetic dielectric transparent in the near infrared spectral range [17]. For photon energy at 0.95 eV the total absorption in the film is about 12%. It means that the density of absorbed photons is 3.6×10$^{19}$ cm$^{-3}$ and a single photon switches the magnetization in 28 nm$^3$ of the garnet.

**Experimental technique for visualization of photo-magnetic switching**

To investigate the magnetization switching induced by femtosecond linearly polarized laser pulses in YIG:Co thin film, we employed the technique of magneto-optical polarized microscopy in Faraday geometry. The magnetic contrast in polarizing microscope comes from the fact that magnetic domains with different out-of-plane magnetization component will result in different angles of the Faraday rotation. The light emitted from the LED source, passing through the polarizer will acquire different polarization rotation in different magnetic domains. The differences can be visualized with the help of an analyzer and a CCD camera. The obtained images visualize magnetic domains. The garnet film was excited with one or and multiple (*N*) pulses. The duration of each pulse was about 50 fs and pulse-to-pulse separation $t_R$ was varied. The pump beam with the fluence below 250 mJ cm$^{-2}$ was focused to a spot about 130 μm in diameter. The wavelength of the pump pulses was tuned within the range between 1000 nm and 1500 nm. The images of magnetic domains were taken before and after the pump excitation. Taking the difference between the images we deduced photo-induced effects on the magnetization in the garnet. All measurements were done in zero applied magnetic field and at room temperature.

Figure S1 (see extended data) shows how the switching depends on pulse-to-pulse separation $t_R$ in the experiment. The switching was observed for single pump and multiple-pulse (*N*=5) excitation with pulse-to-pulse separation time in the range from 1 ms to 20 ms. Whereas for pump pulses at λ=1300 nm there is no visible accumulation effect i.e. the switched domain area does not depend on $t_R$, accumulation is clearly important for the case of λ=1140 nm. An increase of $t_R$ results in a decrease of the switched area. It can be explained by a movement of the domain walls due to magnetostatic effects that typically occurs on the millisecond time scale [21].

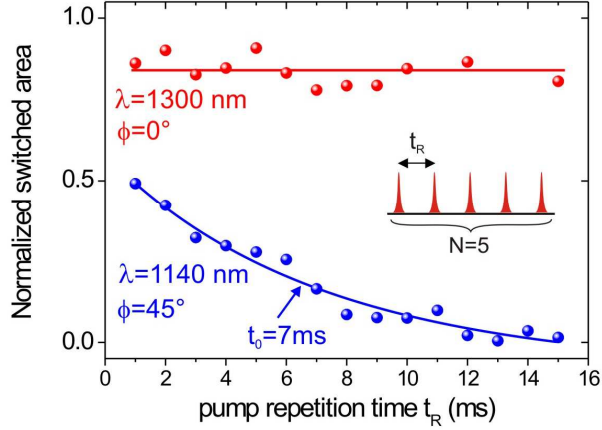

**Extended Data Figure S1 | Photo-magnetic recording in YIG:Co with delay between pump pulses.** The study was performed by means of analysis of normalized switched area with N=5 laser pump pulses as a function of the pump pulse-to-pulse separation time $t_R$. The blue solid line was fitted using exponential function with the characteristic time $t_0$=7 ms.

**Theoretical model of photo-magnetic switching**

The expression for the photo-magnetic tensor in Voigt notation is:

$$\hat{\beta}_V = \begin{pmatrix} b_{1111} & b_{1122} & b_{1133} & 0 & 0 & 0 & 0 & 0 & 0 \\ b_{1122} & b_{1111} & b_{1133} & 0 & 0 & 0 & 0 & 0 & 0 \\ b_{3311} & b_{3311} & b_{3333} & 0 & 0 & 0 & 0 & 0 & 0 \\ 0 & 0 & 0 & b_{2323} & 0 & 0 & b_{2332} & 0 & 0 \\ 0 & 0 & 0 & 0 & b_{3131} & 0 & 0 & b_{3131} & 0 \\ 0 & 0 & 0 & 0 & 0 & b_{1212} & 0 & 0 & b_{1221} \\ 0 & 0 & 0 & b_{3131} & 0 & 0 & b_{3131} & 0 & 0 \\ 0 & 0 & 0 & 0 & b_{2323} & 0 & 0 & b_{2323} & 0 \\ 0 & 0 & 0 & 0 & 0 & b_{1221} & 0 & 0 & b_{1221} \end{pmatrix} \quad (3)$$

We take the action of the laser pulse as a perturbation to the initial state of the magnetization in the film, obtained from the LLG-equation Eq. (1). Depending on the axis of linear polarization of light it gives:

$$E \parallel [100]([010]): \begin{cases} \theta(0) = \theta_0 \pm A \dfrac{1}{1+\alpha^2}\left(a\sin(2\varphi_0)\sin\theta_0 \mp \dfrac{\alpha}{2}b\sin(2\varphi_0)\right) \\ \varphi(0) = \varphi_0 + A \dfrac{1}{1+\alpha^2}\left(b\cos\theta_0 \pm \alpha a\sin(2\varphi_0)\right), \end{cases}$$

$$E \parallel [1\pm10]: \begin{cases} \theta(0) = \theta_0 - \dfrac{1}{2}A\dfrac{\alpha}{1+\alpha^2}\left(b\pm c\sin(2\varphi_0)\right)\sin(2\theta_0) \\ \varphi(0) = \varphi_0 + A\dfrac{1}{1+\alpha^2}\left(b\pm c\sin(2\varphi_0)\right)\cos\theta_0, \end{cases}$$

(4)

where $\varphi_0$, $\theta_0$ denote the ground state angles of the initial domain, $a=a(\lambda)$, $b=b(\lambda)$, and $c=c(\lambda)$ are the parameters, which can be expressed through photo-magnetic tensor coefficients and $A=\gamma M_s \langle E^2 \rangle \Delta t$, where $\langle E^2 \rangle$ is the average intensity of the field in the pulse and $\Delta t \approx 20$ ps is the characteristic time of the emergence of the magnetic domain taken from the experiment [10]. After time interval of the order of $\Delta t$ due to relaxation of the photo-excited Co electrons, the magnetization dynamics can be regarded as free (see Fig. 1). As it is seen from Eq. (4), the initial perturbation strongly depends on the Gilbert damping, initial state and the intensity of light; these are parameters which are crucial for the switching. Wavelength dependence of the efficiency of switching implies wavelength dependence of the photo-magnetic tensor coefficients, namely $a(\lambda)$, $b(\lambda)$, and $c(\lambda)$.